\documentclass[3p]{elsarticle}
\usepackage{amsmath}
\usepackage{subcaption}
\usepackage{graphicx}
\usepackage{tikz}
\usepackage{wasysym}

\usepackage[colorlinks=true]{hyperref}
\usepackage{cleveref}

\DeclareMathOperator{\sign}{\mathrm{sign}}

\begin{document}

\begin{frontmatter}
\title{Heider balance of a chain of actors as dependent on the interaction range and a thermal noise}

\author{Krzysztof Malarz\corref{km}\fnref{orcidkm}}  
\ead{malarz@agh.edu.pl}
\author{Krzysztof Ku{\l}akowski\fnref{orcidkk}}
\ead{kulakowski@fis.agh.edu.pl}

\address{AGH University of Science and Technology,
Faculty of Physics and Applied Computer Science,\\
al. Mickiewicza 30, 30-059 Kraków, Poland.}

\date{August 14, 2020}

\cortext[km]{Corresponding author}
\fntext[orcidkm]{\includegraphics[scale=0.1]{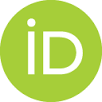} \href{http://orcid.org/0000-0001-9980-0363}{0000-0001-9980-0363}}
\fntext[orcidkk]{\includegraphics[scale=0.1]{ORCID} \href{http://orcid.org/0000-0003-1168-7883}{0000-0003-1168-7883}}

\begin{abstract}
The Heider balance is investigated in a chain of actors, with periodic boundary conditions
and the neighborhood of range $r$, with $r$ as a parameter. 
Two model dynamics are applied: a deterministic cellular automaton (Malarz et al, Physica D 411 (2020) 132556)
and the heat-bath algorithm, with the density of unbalanced-balanced triads in the role of energy.
The outcome is a spectrum of energy in stationary and blinking states and a balanced-unbalanced network transition driven by thermal noise. The critical point $T_c$  increases with the range $r$ and it does not depend on the system size.
\end{abstract}


\end{frontmatter}

\section{Introduction}

The concept of structural balance (Heider balance, HB) has been born in social psychology \cite{Heider,Festinger,Abelson_1968}, it remains basic in mathematical sociology 
\cite{Bonacich}, and it adds \cite{Saito_2011,Nray_2016} to the so-called relational perspective in sociology \cite{Emirbayer_1997}.  Here we focus on its computational aspects when 
applied to networks \cite{Cartwright}. For these purposes HB can be and often is presented as related to interpersonal relations in triads. 
A relation between two individuals can be friendly (positive) or hostile (negative). In a triad of persons A, B and C, represented by vertices of a triangle, there are 
three relations: AB, BC and CA. The triad is balanced if the following rules are true \cite{Aronson}:
\begin{enumerate}
\item my friend's friend is my friend;
\item my enemy's friend is my enemy;
\item my friend's enemy is my enemy;
\item my enemy's enemy is my friend.
\end{enumerate}

This means that in triad in a balanced state either there are two negative relations or all relations are positive. In other words, the product of three relations 
is positive. If all triads in a complete graph are balanced then two parts of the graph appear, where relations within each part are positive and relations between vertices in different parts are negative \cite{Cartwright} or a paradise of only positive relations occurs.

Until recently, most computational effort has been devoted to the case of a complete graph. In Ref.~\cite{Antal}, a discrete algorithm (the so-called Constrained Triad Dynamics) has been designed as to 
lead to HB in complete graphs. In Refs.~\cite{Kulakowski2005,Marvel_2011}, a set of differential equations has been proposed with the same purpose. In both approaches, the time evolution 
of the relations has been shown to end up in some jammed (not balanced) states. To detect them, a work function $U$ has been used, defined as
\begin{equation}
\label{eq:Udef}
U\equiv-\frac{1}{\mathcal{N}}\sum_{ABC} x_{AB} x_{BC} x_{CA},
\end{equation}
where $\mathcal{N}$ is the number of triads, and $x_{ij}$ is the relation between vertices $i$ and $j$. For balanced states of network $U=-1$. Jammed states produce positive deviations from this value, both for symmetric \cite{Antal,Stojkow_2020} and asymmetric \cite{Hassanibesheli_2017a,Krawczyk,Krawczyk_2015} relations. 
On the other hand, treating $U$ as energy allowed to develop a thermodynamic formalism and to identify a phase transition to HB in low temperatures \cite{PhysRevE.99.062302,PhysRevE.100.022303,1911.13048}.

The rationale for using complete graphs is that once a graph is connected, the four rules given above allow to create links between each pair of vertices. 
Yet it is of interest to limit the number of links by applying the time evolution to a network with given structure.
This is equivalent to an assumption that some relations remain absent, what is realistic for larger systems \cite{Dunbar_1992}.
On the other hand, it opens a question how the balance depends on the network topology.
Up to our knowledge, first computational step on this path has been made in Ref.~\cite{2005.11402}, where partial HB ($-1<U<0$) has been identified in a triangular lattice.
There, some local configurations of links have been shown to blink periodically, with period two.
This characteristics was due to the deterministic and synchronous character of time evolution \cite{2005.11402}.
On the other hand, even partially balanced states have been shown to disappear in the triangular lattice if small thermal noise is added \cite{2007.02128}.
On the contrary, HB in complete graphs have been demonstrated to vanish in a finite temperature $T_c$, which increased with the number of vertices \cite{1911.13048}.
Summarizing, interesting phenomena can appear at the meeting of topology and noise.

\begin{figure}
\begin{subfigure}{0.99\textwidth}
\caption{\label{fig:WSr1}$M=7$, $r=1$}
\begin{tikzpicture}[scale=.8]
\foreach \x  in {0,1,2,3,4,5,6}
    \node[draw,fill,circle,scale=.6] at (\x,0) {};
\foreach \x  in {0,1,2,3,4,5,6}
    \node[draw,fill,circle,scale=.6] at (\x,0) {};
    
\draw[thick] (0,0)--(6,0);
\draw[dashed] (-4,0)--(0,0);
\draw[dashed] ( 6,0)--(10,0);
\draw[thick] (11,0)--(12,0);

\node[draw,fill,circle,scale=.6] at (11,0) {};
\node[left] at  (11,0) {$i$};
\node[draw,fill,circle,scale=.6] at (12,0) {};
\node[right] at (12,0) {$i+1=i+r$};
\end{tikzpicture}
\end{subfigure}
\begin{subfigure}{0.99\textwidth}
\caption{\label{fig:WSr2}$M=7$, $r=2$}
\begin{tikzpicture}[scale=.8]
\foreach \x  in {0,1,2,3,4,5,6}
    \node[draw,fill,circle,scale=.6] at (\x,0) {};
\foreach \x  in {2,3,4,5,6}
    \draw[thick]  (\x,0) arc (0:180:1);
\foreach \x  in {0,1,7,8}    
    \draw[dashed] (\x,0) arc (0:180:1);

\draw[thick]   (0,0)--(6,0);
\draw[dashed] (-4,0)--(0,0);
\draw[dashed] ( 6,0)--(10,0);
\draw[thick]  (11,0)--(13,0);

\node[draw,fill,circle,scale=.6] at (11,0) {};
\node[left] at (11,0) {$i$};
\node[draw,fill,circle,scale=.6] at (12,0) {};
\node[above] at (12,0) {$i+1$};
\node[draw,fill,circle,scale=.6] at (13,0) {};
\node[right] at (13,0) {$i+2=i+r$};
\draw[thick] (13,0) arc (0:180:1);
\end{tikzpicture}
\end{subfigure}
\begin{subfigure}{0.99\textwidth}
\caption{\label{fig:WSr3}$M=7$, $r=3$}
\begin{tikzpicture}[scale=.8]
\foreach \x  in {0,1,2,3,4,5,6}
    \node[draw,fill,circle,scale=.6] at (\x,0) {};
\foreach \x  in {2,3,4,5,6}
    \draw[thick]  (\x,0) arc (0:180:1);
\foreach \x  in {0,1,7,8}    
    \draw[dashed] (\x,0) arc (0:180:1);

\foreach \x  in {3,4,5,6}
    \draw[thick] (\x,0) arc (0:180:1.5);

\foreach \x  in {0,1,2,3,7,8,9}
    \draw[dashed] (\x,0) arc (0:180:1.5);
    
\draw[thick] (0,0)--(6,0);
\draw[dashed] (-4,0)--(0,0);
\draw[dashed] ( 6,0)--(10,0);
\draw[thick] (11,0)--(14,0);

\node[draw,fill,circle,scale=.6] at (11,0) {};
\node[left] at (11,0) {$i$};
\node[draw,fill,circle,scale=.6] at (12,0) {};
\node[above] at (12,0) {$i+1$};
\node[draw,fill,circle,scale=.6] at (13,0) {};
\node[below] at (13,0) {$i+2$};
\node[draw,fill,circle,scale=.6] at (14,0) {};
\node[right] at (14,0) {$i+3=i+r$};

\draw[thick] (13,0) arc (0:180:1);
\draw[thick] (14,0) arc (0:180:1);
\draw[thick] (14,0) arc (0:180:1.5);
\end{tikzpicture}
\end{subfigure}
\caption{\label{fig:WS}Network construction scheme.}
\end{figure}
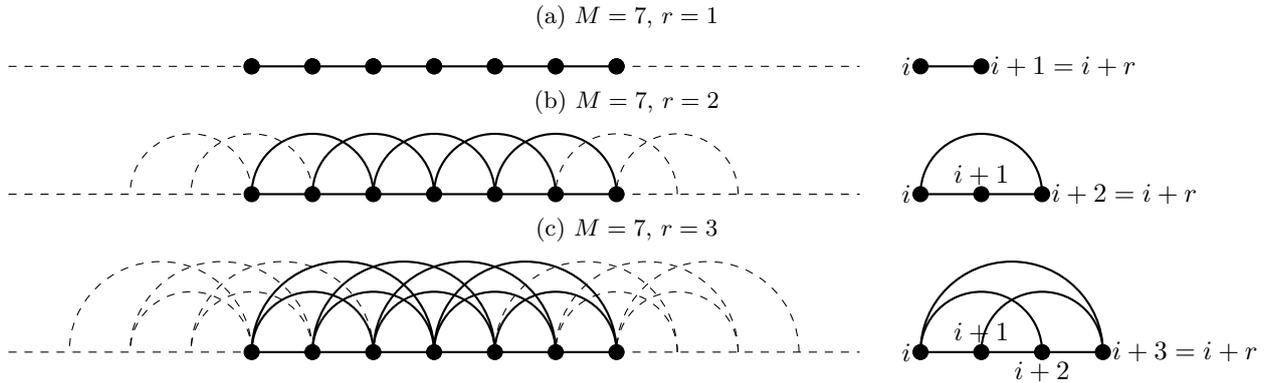

Our aim here is to investigate the structure of linear chain of actors, with the range $r$ of interaction between them as a parameter. Such a structure is usually used as an initial form of the  Watts--Strogatz network \cite{Watts-Strogatz}; here the rewiring is postponed. 
For given $r$, the number of links $Mr$ increases only linearly with the system size $M$, and the mean degree of vertices remain constant ($2r$), on the contrary to the complete graphs.
However, when $r$ increases, the system is driven towards the complete graph.
Further, `thermal' noise \cite{0807.0711} is added via the heat-bath algorithm \cite{Binder_1997,ISI:A1986C512100020,ISI:000076966100035}.
As we see, the system is versatile enough to link some previous approaches \cite{PhysRevE.99.062302,PhysRevE.100.022303,1911.13048,2005.11402,2007.02128}.
Last but not least, we are not aware of any research of HB in the chain structure.

In this paper, we are going to explore the system behavior with two methods.
First is the deterministic synchronous cellular automaton without noise \cite{2005.11402}. 
For small systems, it is possible to check the final state for all ($2^{Mr}$) initial states of the network.
For larger systems we rely on random sampling of initial states.
If---by chance---an initial state is balanced, it remains unchanged during the time evolution, and $U=-1$.
We are interested if the system ends in a balanced state also for other initial states.
Second method is the heat-bath algorithm, as in Refs.~\cite{1911.13048,2007.02128}.
We ask if a phase transition between balanced and unbalanced state is observed and---if so---how it depends on the interaction range $r$ and the system size $M$. 

Next three Sections are devoted to calculations, results and discussion, respectively. On the contrary to our previous works \cite{Stojkow_2020,Krawczyk}, 
here we skip the discussion of social aspects of the model. In a sense, the Heider dynamics is flying on its own. Possible applications beyond social psychology are mentioned at the end of the text.

\section{Methods}
To write the work function $U$, we need to distinguish triads, not to omit any and not to include any of them twice. A triad is equivalent to three vertices, which are discrete points on a line (see \Cref{fig:WS}). It is convenient to assign to each triad the position $i=0,\cdots,M-1$ of its left end and the length $j$ of its longest edge $(2\leq j\leq r)$. If $M>3r$, the position $k$ of the third vertex is between the former two. Then the work function \eqref{eq:Udef} is
\begin{equation}
\label{eq:U}
  U(M,r)= -\dfrac{2}{M(r-1)r} \sum_{i=0}^{M-1} \sum_{j=2}^r \sum_{k=1}^{j-1} x_{i,[i+j]_M}\cdot x_{[i+j]_M,[i+j-k]_M}\cdot x_{[i+j-k]_M,i},
\end{equation}
where $[\cdots]_M$ stands for modulo $M$ function: $[a]_M\equiv a-M\cdot\lfloor a/M\rfloor$, to keep the periodic boundary conditions. Links values are updated synchronously.

\subsection{\label{sec:deterministic}Deterministic evolution, $T=0$}

In absence of thermal noise the system evolution may be realised with synchronous cellular automaton with the rule of links update as
\begin{subequations}
\label{eq:xevolT0}
\begin{equation}
x'_{i,[i+k]_M}=x'_{[i+k]_M,i}=
\begin{cases}
 \sign(\xi_{i,[i+k]_M}) & \text{ if } 1\le k\le r \text{ and } \xi_{i,j}\ne 0,\\
 x_{i,[i+k]_M}          & \text{ if } 1\le k\le r \text{ and } \xi_{i,j}=0,\\
 0                      & \text{ otherwise},
\end{cases}
\end{equation}
where $x'_{i,j}\equiv x_{i,j}(t+1)$, $x_{i,j}\equiv x_{i,j}(t)$, $x_{i,j}=0$ means that link between actors $i$ and $j$ does not exist and $\xi_{i,j}\equiv \xi_{i,j}(t)$ is defined as 
\begin{equation}
\label{eq:xi}
\xi_{i,[i+k]_M}=\xi_{[i+k]_M,i}=
    \begin{cases}
    \sum_{m=i+k-r}^{i+r} x_{i,[m]_M}\cdot x_{[m]_M,[i+k]_M} & \iff 1\le k\le r,\\
    0 & \iff k=0 \text{ or } r<k<M-r.
    \end{cases}
\end{equation}
\end{subequations}

In these and earlier \cite{2005.11402} simulations, a state of a link was modified during the time evolution only if keeping all other links unchanged, the energy decreased. This assumption is chosen to keep the coherence with the aspect of human action, as discussed by Herbert Simon \cite{Kaufman_1999}: decisions are `satisficing' rather than optimal. In this sense a state of a link is accepted if its change---keeping other links constant---does not reduce the energy of the system.

\begin{figure}[htbp]
\centering
\includegraphics[width=.35\textwidth]{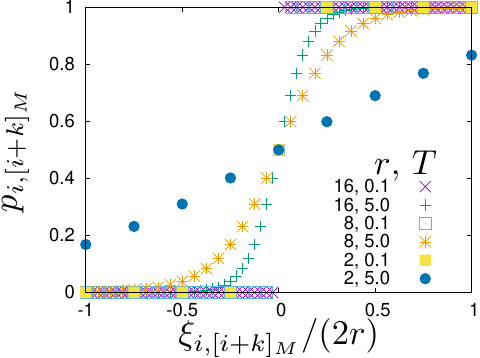}
\caption{\label{fig:pvsxi}Probability $p$ of setting-up link value to the value of $+1$ as dependent on $\xi$ parameter for various $r$ and $T$. For low temperatures $T$ the dependence $p(\xi)$ is the Heaviside step function except of $\xi=0$, where $p=1/2$.}
\end{figure}

\subsection{Stochastic evolution, $T>0$}

In order to include the effect of the thermal noise we utilise the heat-bath algorithm \cite{Binder_1997,ISI:A1986C512100020,ISI:000076966100035}:
\begin{subequations}
\label{eq:xevolTT}
\begin{equation}
\label{eq:xi_TT}
x'_{i,[i+k]_M}=x'_{[i+k]_M,i}=
    \begin{cases}
	+1 & \text{ with probability }p_{i,[i+k]_M} \text{ for } 1\le k\le r,\\
	-1 & \text{ with probability }(1-p_{i,[i+k]_M})  \text{ for } 1\le k\le r,\\
	 0 & \text{ for } k=0 \text{ and } r<k<M-r 
    \end{cases}
\end{equation}
and
\begin{equation}
    \label{eq:prob}
    p_{i,[i+k]_M}=\frac{\exp(\xi_{i,[i+k]_M}/T)}{\exp(\xi_{i,[i+k]_M}/T)+\exp(-\xi_{i,[i+k]_M}/T)},
\end{equation}
where $\xi_{i,[i+k]_M}$ are given in \Cref{eq:xi} and $T$ plays a role of social temperature \cite{0807.0711}.
\end{subequations}

The dependence of $p_{i,[i+k]_M}$ on $\xi_{i,[i+k]_M}$ (normalised to the double range of interaction) for various ranges of interaction $r$ and various values of temperatures $T$ are presented in \Cref{fig:pvsxi}. Please note, that for low temperatures (e.g. $T=0.1$) formula \eqref{eq:xevolTT} reduces to deterministic case \eqref{eq:xevolT0} except of $\xi=0$, where $p=1/2$. In the latter case the link value $x'_{i,[i+k]_M}=x'_{[i+k]_M,i}=\pm 1$ is taken randomly, while in deterministic case $x'_{i,[i+k]_M}=x'_{[i+k]_M,i}=x_{i,[i+k]_M}$ does not evolve.

\section{Results}

\subsection{Deterministic evolution, $T=0$}

\subsubsection{Brute force results}

For relatively small systems, it is possible to check all possible initial configurations of the relations (brute force). For computational reasons, we have applied this method 
mostly to the case $r=2$.
There, for example for $M=7$ we have $2^{Mr}=2^{14}=$16,384 available configurations of the links. There is only $2^M=128$ initially balanced states among them, yet---as shown in \Cref{tab:bruteforce}---there is almost 62 percent of initial states which lead to the balance. The remaining 38 percent of initial states lead to blinking states, with period two. 
However, for larger systems the percentage of blinking states increases. On the other hand, for given $r$ the average work function $\langle U\rangle$ is equal to $-0.75$ for $r=2$, and about $-0.90$ for $r=3$. The latter value is obtained for $M=10$ (30 links, $2^{30}$ states). 

\begin{table}[tb]
    \caption{\label{tab:bruteforce} Brute force results. The second column indicates the number ($Mr$) of links in the system. $2^M$ among $2^{Mr}$ states are balanced already at $t=0$. Simulation takes $t_{\max}=50$ time steps. Brackets $\langle\cdots\rangle$ stand for averaging over $10^7$ initial states. The last three columns present fractions of finally balanced, fixed and blinking states, respectively.}
    \centering
    \begin{tabular}{rrrrrrrr}
    \hline\hline
        &    &         number &                   & initially& \multicolumn{3}{c}{finally} \\ 
        \cline{6-8}
    $M$ & $L$& of states& $\langle U\rangle$& balanced & balanced & fixed & blinking\\ \hline
    \multicolumn{8}{c}{$r=2$}\\ \hline
     7& 14&        16,384&    $-0.75$&    128& 61.72\% & 61.72\% & 38.28\%\\
     8& 16&        65,536&    $-0.75$&    256& 57.81\% & 57.81\% & 42.19\%\\
     9& 18&       262,144&    $-0.75$&    512& 53.71\% & 53.71\% & 46.29\%\\
    10& 20&     1,048,576&    $-0.75$&  1,024& 50.20\% & 50.20\% & 49.80\%\\
    11& 22&     4,194,304&    $-0.75$&  2,048& 46.83\% & 46.83\% & 53.17\%\\
    12& 24&    16,777,216&    $-0.75$&  4,096& 43.72\% & 43.73\% & 56.27\%\\
    13& 26&    67,108,864&    $-0.75$&  8,192& 40.81\% & 40.81\% & 59.19\%\\
    14& 28&   268,435,456&    $-0.75$& 16,384& 38.09\% & 38.09\% & 61.91\%\\
    15& 30& 1,073,741,824&    $-0.75$& 32,768& 35.55\% & 35.55\% & 64.45\%\\
    \hline
    \multicolumn{8}{c}{$r=3$}\\ \hline
    10& 30& 1,073,741,824& $-0.90355$&  1,024& 65.75\% & 90.99\% &  9.01\%\\
    \hline\hline
    \end{tabular}
   \end{table}

\begin{figure}[htbp]
\centering
\begin{subfigure}{0.31\textwidth}
\centering
\caption{$M=7$, $\langle U\rangle=-0.75$}
\includegraphics[width=.99\textwidth]{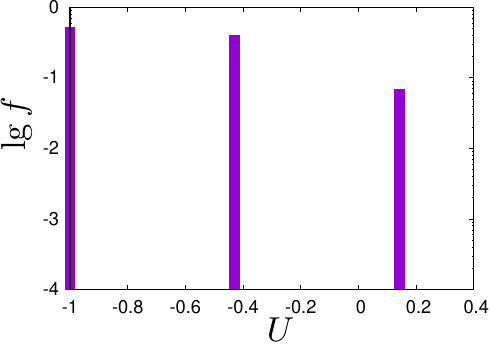}
\end{subfigure}
\hfill 
\begin{subfigure}{0.31\textwidth}
\centering
\caption{$M=11$, $\langle U\rangle=-0.75$}
\includegraphics[width=.99\textwidth]{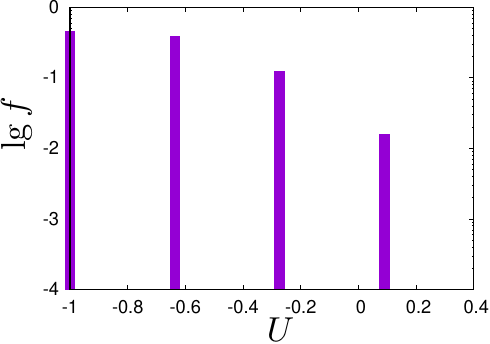}
\end{subfigure}
\hfill 
\begin{subfigure}{0.31\textwidth}
\centering
\caption{$M=15$, $\langle U\rangle=-0.75$}
\includegraphics[width=.99\textwidth]{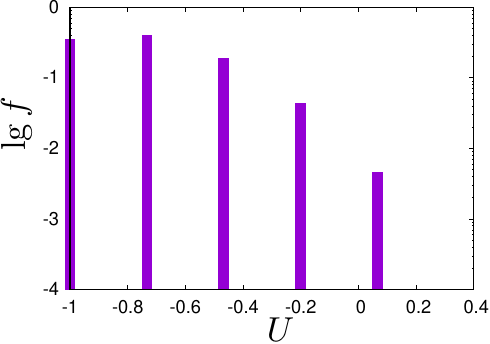}
\end{subfigure}
\caption{\label{fig:r2histobrut}Probability distribution function $f$ of the final system work function $U$ for $r=2$. Brute force results.}
\end{figure}

\begin{table}[htbp]
    \caption{\label{tab:probab}Sampling $10^7$ among $2^{Mr}$ initial states. The second column indicates the number ($Mr$) of links in the system. Brackets $\langle\cdots\rangle$ stand for averaging over $10^7$ initial states. Simulation takes $t_{\max}=50$ time steps. The last three columns present fractions of finally balanced, fixed and blinking states, respectively. For $r>2$ some jammed states appear, neither balanced nor blinking, but their frequency decreases with $M$.}
    \centering
    \begin{tabular}{rrrrrr}
    \hline\hline
        &    &                  & \multicolumn{3}{c}{finally} \\ 
        \cline{4-6}
    $M$ & $L$& $\langle U\rangle$& balanced & fixed & blinking\\ \hline
    \multicolumn{6}{c}{$r=2$}\\ \hline
      20 &   40 & $-0.74995$ &  25.17\% &  25.17\% &   74.83\%\\
      30 &   60 & $-0.74991$ &  12.64\% &  12.64\% &   87.36\%\\
      40 &   80 & $-0.74995$ &   6.35\% &   6.35\% &   93.65\%\\
      50 &  100 & $-0.74994$ &   3.19\% &   3.19\% &   96.81\%\\
     100 &  200 & $-0.75001$ &   0.10\% &   0.10\% &   99.90\%\\
     200 &  400 & $-0.75001$ & 0.0001\% & 0.0001\% & 99.9999\%\\
     300 &  600 & $-0.75000$ &      0\% &      0\% &     100\%\\
     500 & 1000 & $-0.74999$ &      0\% &      0\% &     100\%\\
    1000 & 2000 & $-0.75001$ &      0\% &      0\% &     100\%\\
    \hline
    \multicolumn{6}{c}{$r=3$}\\ \hline
      20 &   60 & $-0.90407$ &  43.32\% & 82.77\% & 17.23\%\\
      30 &   90 & $-0.90406$ &  28.50\% & 75.31\% & 24.69\%\\
      40 &  120 & $-0.90406$ &  18.76\% & 68.52\% & 31.48\%\\
      50 &  150 & $-0.90406$ &  12.33\% & 62.35\% & 37.65\%\\
     100 &  300 & $-0.90404$ &   1.52\% & 38.87\% & 61.13\%\\
     200 &  600 & $-0.90407$ &  0.023\% & 15.11\% & 84.89\%\\
     300 &  900 & $-0.90407$ &0.00033\% &  5.88\% & 94.12\%\\
     500 & 1500 & $-0.90407$ &      0\% &  0.89\% & 99.11\%\\
     700 & 2100 & $-0.90408$ &      0\% &  0.13\% & 99.87\%\\
     \hline
     \multicolumn{6}{c}{$r=4$}\\ \hline
      20 &   80 & $-0.94376$ &  63.04\% & 78.00\% & 22.00\%\\
      30 &  120 & $-0.94378$ &  50.07\% & 68.91\% & 31.09\%\\
      40 &  160 & $-0.94378$ &  39.76\% & 60.86\% & 39.14\%\\
      50 &  200 & $-0.94378$ &  31.57\% & 53.74\% & 46.26\%\\
     100 &  400 & $-0.94376$ &   9.96\% & 28.86\% & 71.14\%\\
     200 &  800 & $-0.94375$ &   0.99\% &  8.34\% & 91.66\%\\ 
     300 & 1200 & $-0.94376$ &  0.099\% &  2.41\% & 97.59\%\\ 
     400 & 1600 & $-0.94375$ & 0.0095\% &  0.70\% & 99.30\%\\ 
     500 & 2000 & $-0.94376$ &0.00095\% &  0.20\% & 99.80\%\\ 
     \hline\hline
    \end{tabular}
\end{table}

\begin{figure}[htbp]
\centering
\begin{subfigure}{0.32\textwidth}
\centering
\caption{$M=10$, $\langle U\rangle\approx-0.90355$}
\includegraphics[width=.99\textwidth]{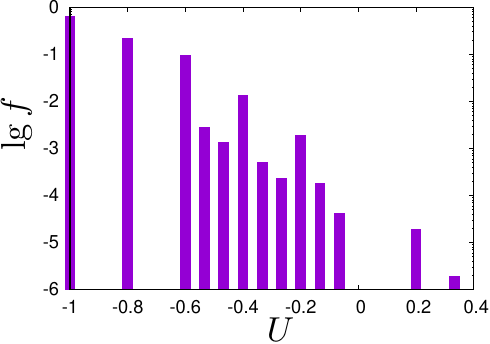}
\end{subfigure}
\caption{\label{fig:r3histobrut}Probability distribution function $f$ of the final system work function $U$ for $r=3$. Brute force results.}
\end{figure}

\subsubsection{Sampling initial states}

\begin{figure}[htbp]
\begin{subfigure}{0.32\textwidth}
\centering
\caption{$M=20$, $\langle U\rangle\approx-0.75$}
\includegraphics[width=.99\textwidth]{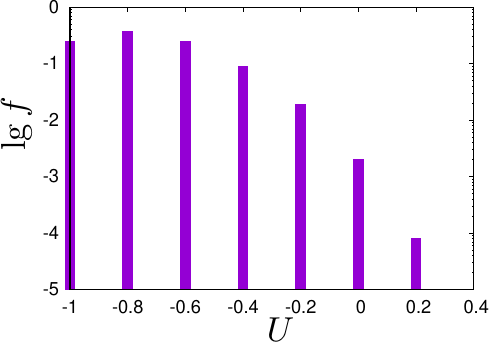}
\end{subfigure}
\hfill 
\begin{subfigure}{0.32\textwidth}
\centering
\caption{$M=30$, $\langle U\rangle\approx-0.75$}
\includegraphics[width=.99\textwidth]{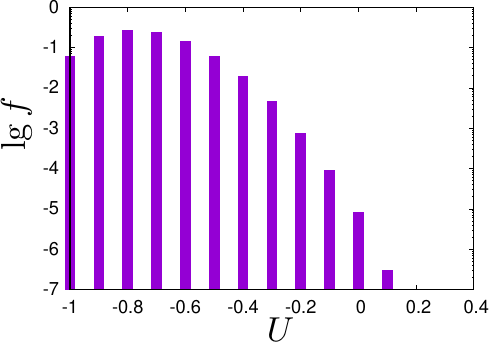}
\end{subfigure}
\hfill 
\begin{subfigure}{0.32\textwidth}
\centering
\caption{$M=100$, $\langle U\rangle\approx-0.75$}
\includegraphics[width=.99\textwidth]{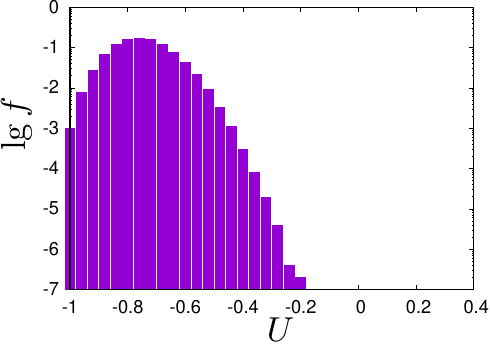}
\end{subfigure}
\hfill 
\begin{subfigure}{0.32\textwidth}
\centering
\caption{$M=200$, $\langle U\rangle\approx-0.75$}
\includegraphics[width=.99\textwidth]{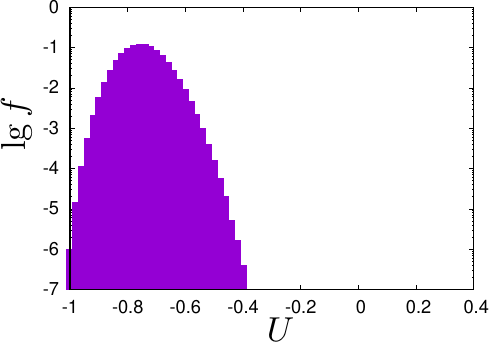}
\end{subfigure}
\hfill 
\begin{subfigure}{0.32\textwidth}
\centering
\caption{$M=500$, $\langle U\rangle\approx-0.75$}
\includegraphics[width=.99\textwidth]{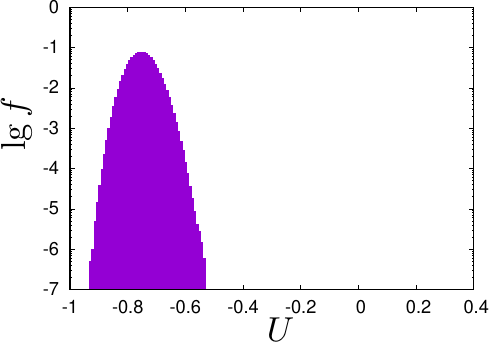}
\end{subfigure}
\hfill 
\begin{subfigure}{0.32\textwidth}
\centering
\caption{$M=1000$, $\langle U\rangle\approx-0.75$}
\includegraphics[width=.99\textwidth]{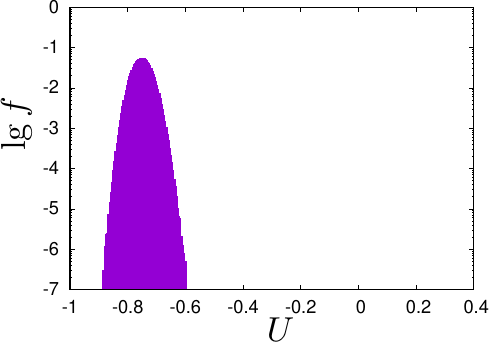}
\end{subfigure}
\caption{\label{fig:r2histoprob}Probability distribution function $f$ of the final system work function $U$ for $r=2$. The results rely on sampling initial state $10^7$ times. For $M\ge 500$ the theoretically predicted probability of occurring (not observed) any of initially balanced state is below $2^M/2^{Mr}\approx 3\cdot 10^{-151}$.}
\end{figure}

\begin{figure}[htbp]
\begin{subfigure}{0.32\textwidth}
\centering
\caption{$M=20$, $\langle U\rangle\approx-0.9040\cdots$}
\includegraphics[width=.99\textwidth]{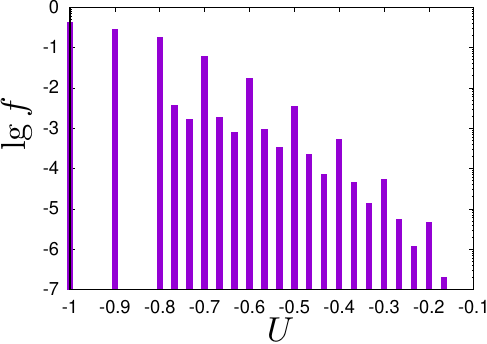}
\end{subfigure}
\hfill 
\begin{subfigure}{0.32\textwidth}
\centering
\caption{$M=40$, $\langle U\rangle\approx-0.9040\cdots$}
\includegraphics[width=.99\textwidth]{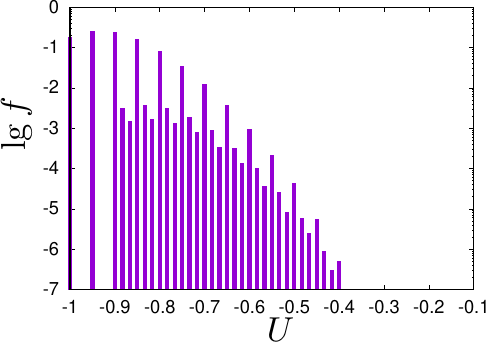}
\end{subfigure}
\hfill 
\begin{subfigure}{0.32\textwidth}
\centering
\caption{$M=50$, $\langle U\rangle\approx-0.9040\cdots$}
\includegraphics[width=.99\textwidth]{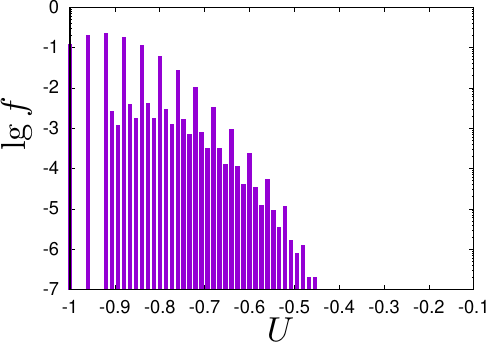}
\end{subfigure}
\hfill 
\begin{subfigure}{0.32\textwidth}
\centering
\caption{$M=200$, $\langle U\rangle\approx-0.9040\cdots$}
\includegraphics[width=.99\textwidth]{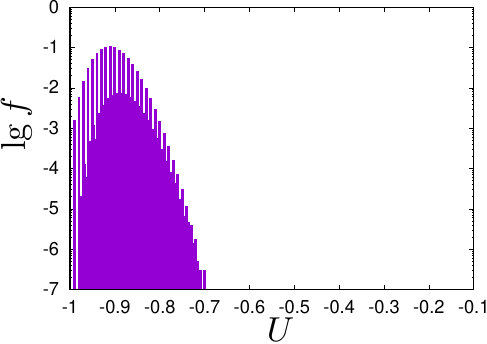}
\end{subfigure}
\hfill 
\begin{subfigure}{0.32\textwidth}
\centering
\caption{$M=500$, $\langle U\rangle\approx-0.9040\cdots$}
\includegraphics[width=.99\textwidth]{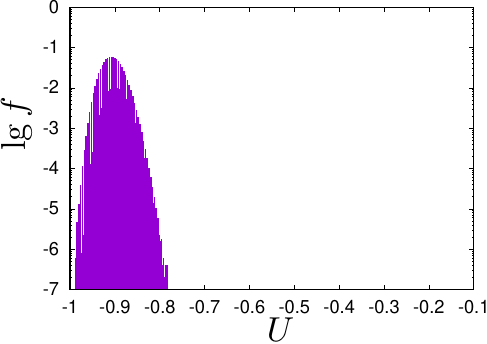}
\end{subfigure}
\hfill 
\begin{subfigure}{0.32\textwidth}
\centering
\caption{$M=700$, $\langle U\rangle\approx-0.9040\cdots$}
\includegraphics[width=.99\textwidth]{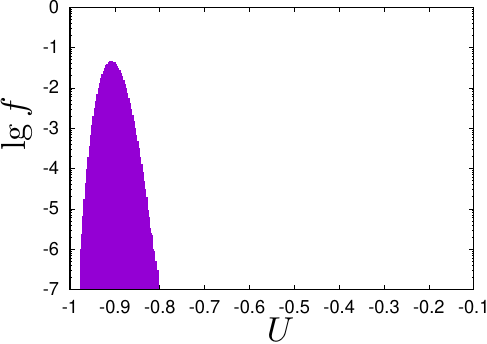}
\end{subfigure}
\caption{\label{fig:r3histoprob}Probability distribution function $f$ of the final system work function $U$ for $r=3$. The results rely on sampling initial state $10^7$ times. For $M\ge 500$ the theoretically predicted probability of occurring (not observed) any of initially balanced state is below $2^M/2^{Mr}\approx 9\cdot 10^{-302}$.}
\end{figure}

\begin{figure}[htbp]
\begin{subfigure}{0.31\textwidth}
\centering
\caption{$M=20$, $\langle U\rangle\approx-0.9437\cdots$}
\includegraphics[width=.99\textwidth]{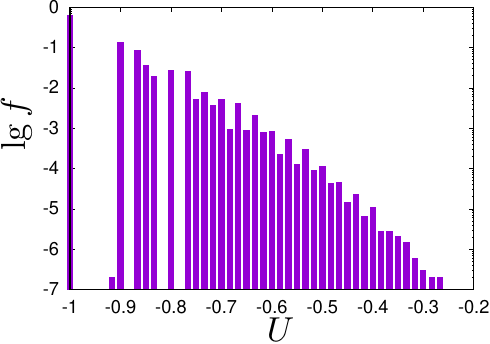}
\end{subfigure}
\hfill 
\begin{subfigure}{0.31\textwidth}
\centering
\caption{$M=40$, $\langle U\rangle\approx-0.9437\cdots$}
\includegraphics[width=.99\textwidth]{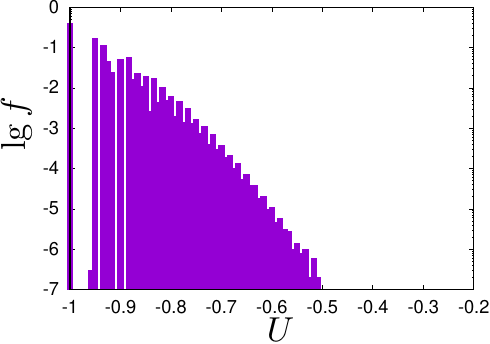}
\end{subfigure}
\hfill 
\begin{subfigure}{0.31\textwidth}
\centering
\caption{$M=50$, $\langle U\rangle\approx-0.9437\cdots$}
\includegraphics[width=.99\textwidth]{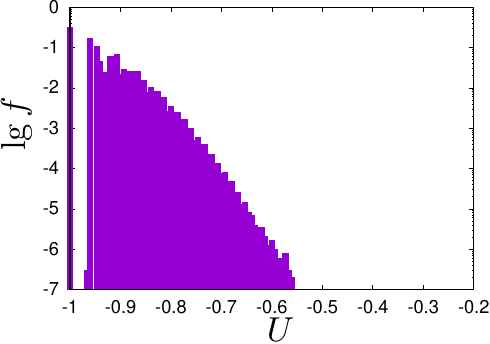}
\end{subfigure}
\hfill 
\begin{subfigure}{0.31\textwidth}
\centering
\caption{$M=100$, $\langle U\rangle\approx-0.9437\cdots$}
\includegraphics[width=.99\textwidth]{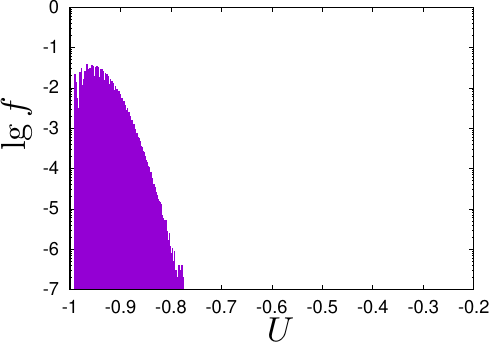}
\end{subfigure}
\hfill 
\begin{subfigure}{0.31\textwidth}
\centering
\caption{$M=200$, $\langle U\rangle\approx-0.9437\cdots$}
\includegraphics[width=.99\textwidth]{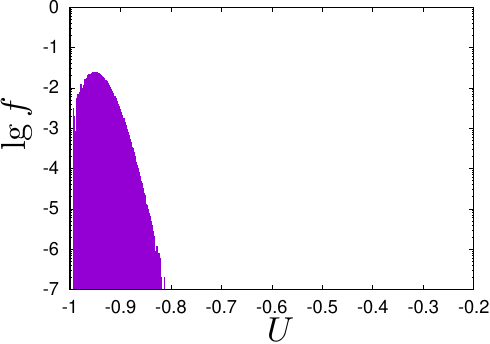}
\end{subfigure}
\hfill 
\begin{subfigure}{0.31\textwidth}
\centering
\caption{$M=500$, $\langle U\rangle\approx-0.9437\cdots$}
\includegraphics[width=.99\textwidth]{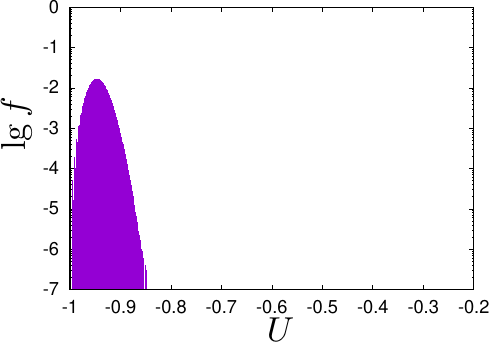}
\end{subfigure}
\caption{\label{fig:r4histoprob}Probability distribution function $f$ of the final system work function $U$ for $r=4$. The results rely on sampling initial state $10^7$ times.}
\end{figure}

For larger systems ($M\ge 20$), we rely on random sampling of $10^7$ initial states. 
The seemingly exact value $-0.75$ (see Ref.~\cite{2005.11402}) of the average work function $\langle U\rangle$ is reached with the accuracy $10^{-4}$ already for $M=20$.
For $r>2$ some jammed states appear, neither balanced nor blinking, but their frequency decreases with $M$.
For $r=3$ the value of the work function $\langle U\rangle=-0.9040\cdots$ is reproduced till $M=700$. For $r=4$ we get $\langle U\rangle=-0.9437\cdots$ (see \Cref{tab:probab}). 
It is remarkable  that $\langle U\rangle$ decreases with $r$ despite the fact that the balanced states (where $U=-1$ is minimal) are less and less probable when $r$ increases.
More insight into this discrepancy is attained when we check the spectra of the work function, shown in \Cref{fig:r2histobrut,fig:r3histobrut,fig:r2histoprob,fig:r3histoprob,fig:r4histoprob}.
These distributions are more and more narrow when $M$ increases, both for $r=2$, 3 and 4. 

\subsection{Stochastic evolution, $T>0$}
Here we report the results including the action of noise.
For convenience and for coherence with literature \cite{PhysRevE.99.062302,PhysRevE.100.022303,1911.13048}
we adopt the formulae of thermal noise; an extension of this interpretation has been suggested in Ref.~\cite{0807.0711}.

\begin{figure}[htbp]
\centering
\begin{subfigure}{0.82\textwidth}
\centering
\caption{$M=256$, $r=2$}
\includegraphics[width=.99\textwidth]{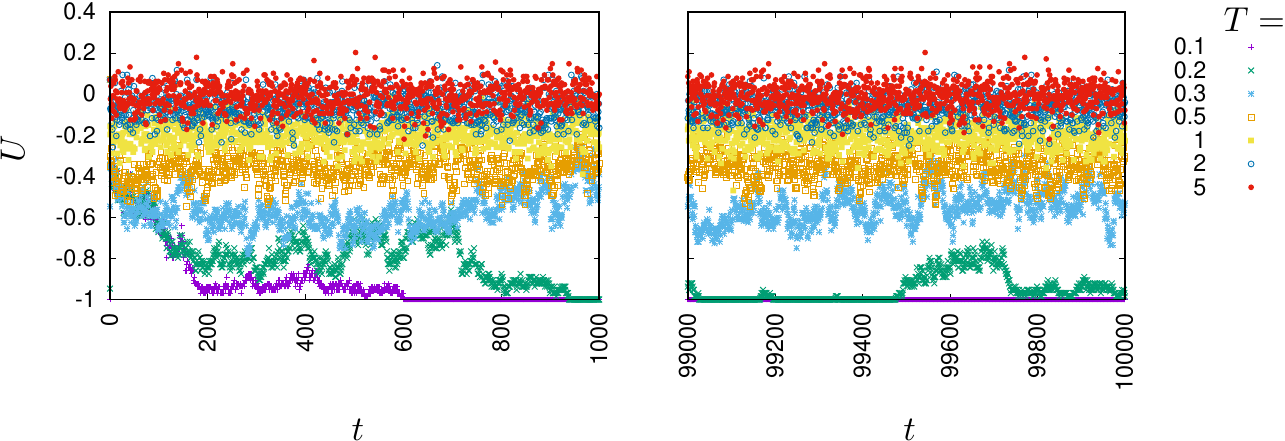}
\end{subfigure}
\begin{subfigure}{0.82\textwidth}
\centering
\caption{$M=512$, $r=2$}
\includegraphics[width=.99\textwidth]{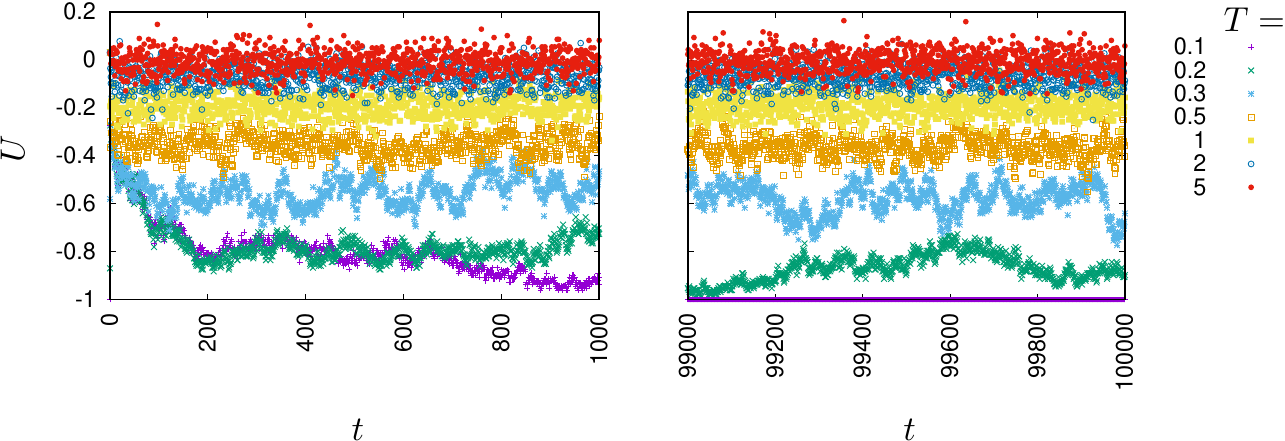}
\end{subfigure}
\begin{subfigure}{0.82\textwidth}
\centering
\caption{$M=256$, $r=12$}
\includegraphics[width=.99\textwidth]{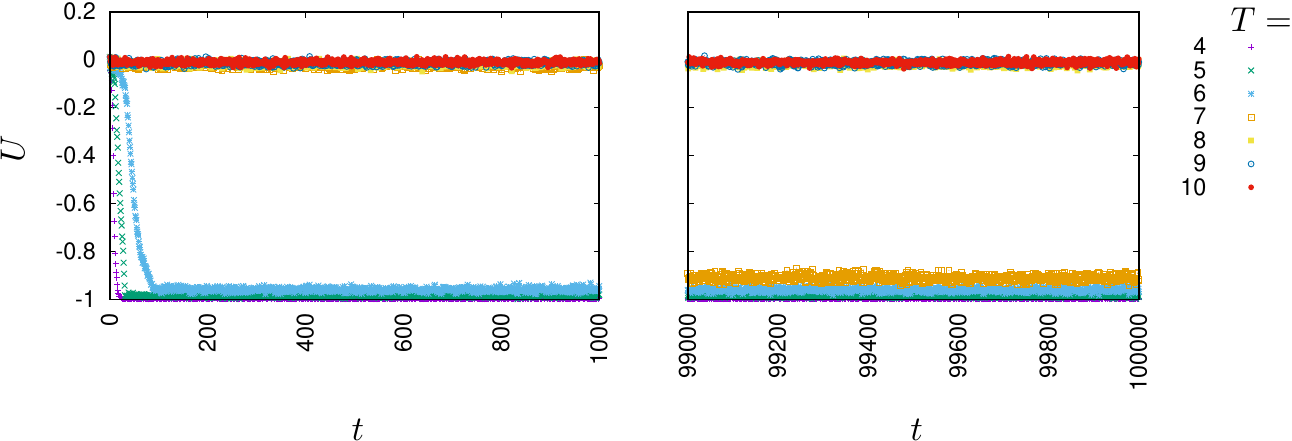}
\end{subfigure}
\begin{subfigure}{0.82\textwidth}
\centering
\caption{$M=512$, $r=12$}
\includegraphics[width=.99\textwidth]{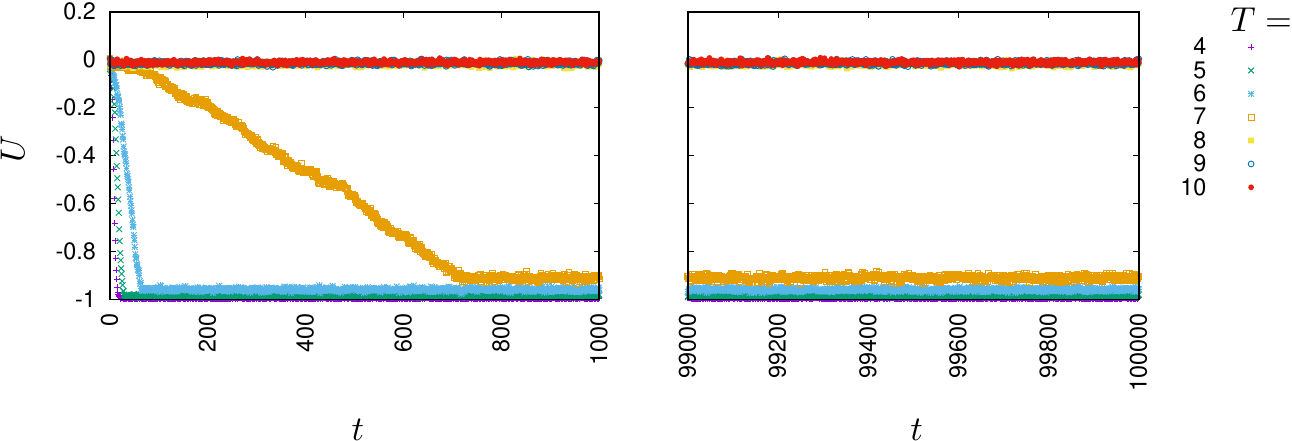}
\end{subfigure}
\caption{\label{fig:Uvst}Time evolution of the system work function $U$ when starting point of simulation is a random state. Single simulation.}
\end{figure}

\begin{figure}[htbp]
\begin{subfigure}{0.31\textwidth}
\caption{$r=2$}
\centering
\includegraphics[width=.99\textwidth]{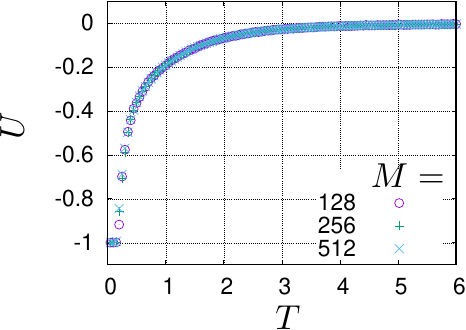}
\end{subfigure}
\begin{subfigure}{0.31\textwidth}
\caption{$r=3$}
\centering
\includegraphics[width=.99\textwidth]{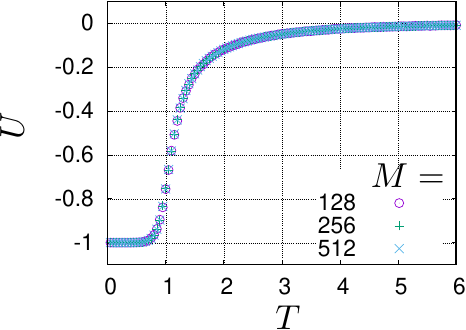}
\end{subfigure}
\begin{subfigure}{0.31\textwidth}
\caption{$r=4$}
\centering
\includegraphics[width=.99\textwidth]{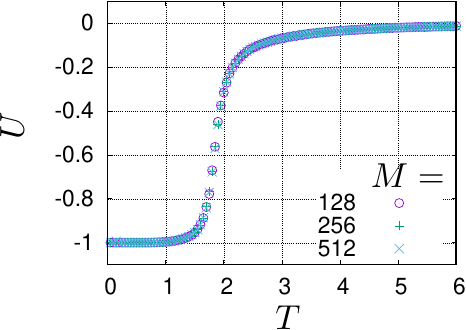}
\end{subfigure}
\begin{subfigure}{0.31\textwidth}
\caption{$r=6$}
\centering
\includegraphics[width=.99\textwidth]{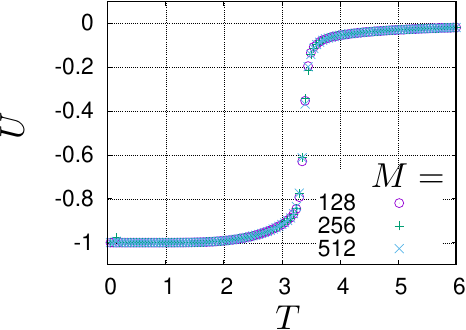}
\end{subfigure}
\begin{subfigure}{0.31\textwidth}
\caption{$r=7$}
\centering
\includegraphics[width=.99\textwidth]{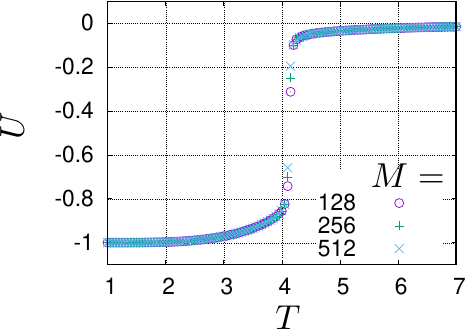}
\end{subfigure}
\begin{subfigure}{0.31\textwidth}
\caption{$r=8$}
\centering
\includegraphics[width=.99\textwidth]{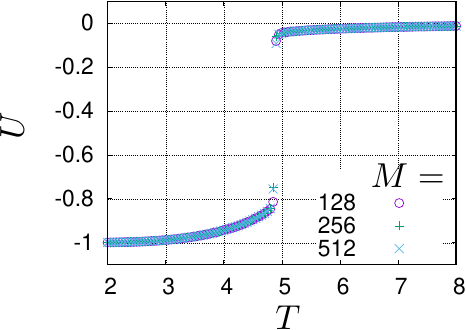}
\end{subfigure}
\hfill 
\begin{subfigure}{0.31\textwidth}
\caption{$r=12$}
\centering
\includegraphics[width=.99\textwidth]{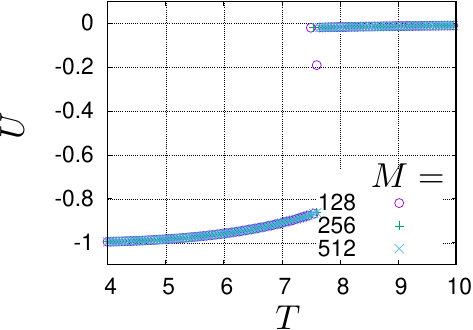}
\end{subfigure}
\hfill 
\begin{subfigure}{0.31\textwidth}
\caption{$r=16$}
\centering
\includegraphics[width=.99\textwidth]{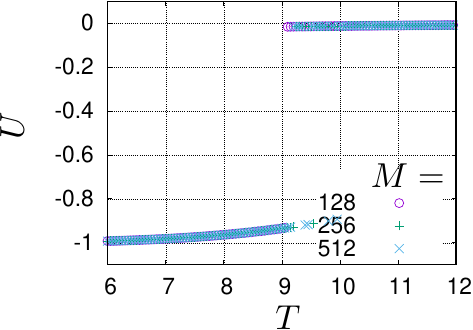}
\end{subfigure}
\hfill 
\begin{subfigure}{0.31\textwidth}
\caption{$r=32$}
\centering
\includegraphics[width=.99\textwidth]{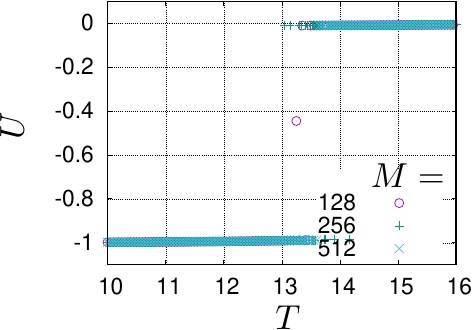}
\end{subfigure}
\caption{\label{fig:UvsT}Thermal evolution of the system work function $U$ when starting point of simulation is a random state. Single simulation. Simulation takes $t_{\max}=10^5$ time steps and $U$ is averaged over the last $\tau=10^4$ time steps.}
\end{figure}

\begin{figure}[htbp]
\begin{subfigure}{0.32\textwidth}
\caption{$M=1024$, $r=64$}
\centering
\includegraphics[width=.99\textwidth]{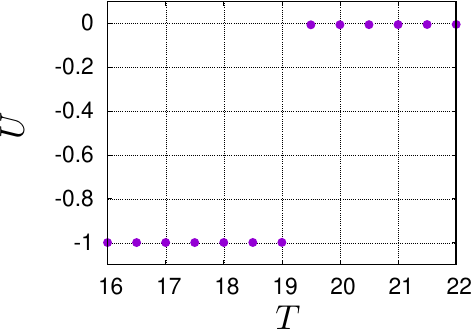}
\end{subfigure}
\hfill 
\begin{subfigure}{0.32\textwidth}
\caption{$M=1024$, $r=128$}
\centering
\includegraphics[width=.99\textwidth]{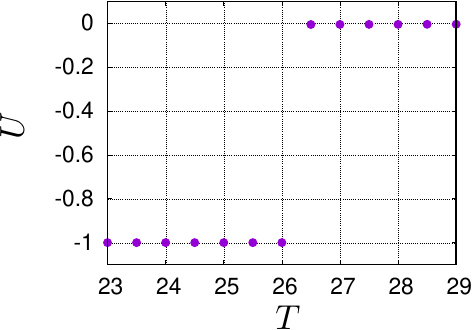}
\end{subfigure}
\hfill 
\begin{subfigure}{0.32\textwidth}
\caption{$M=1024$, $r=256$}
\centering
\includegraphics[width=.99\textwidth]{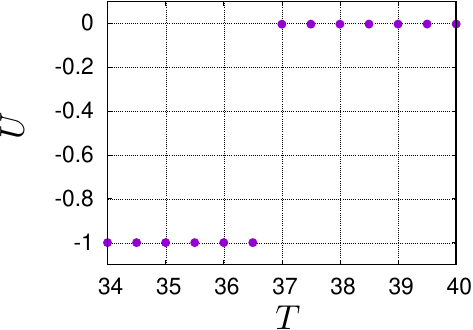}
\end{subfigure}
\caption{\label{fig:UvsT_M=1024}Thermal evolution of the system work function $U$ when starting point of simulation is random state. Single simulation. Simulation takes $t_{\max}=10^5$ time steps and $U$ is averaged over the last $\tau=10^4$ time steps.}
\end{figure}

\begin{figure}[htbp]
    \centering
    \includegraphics[width=0.32\textwidth]{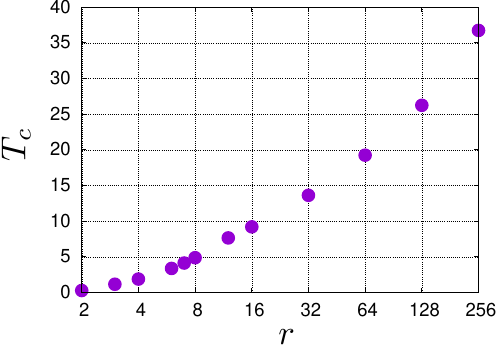}
    \caption{\label{fig:Tcvsr}The critical temperature $T_c$ vs. the range $r$ of interaction. The former is read from the plots $U(T)$ as the value of $T$ where $U=-0.5$.}
\end{figure}

In \Cref{fig:Uvst} we show a series of plots of $U$ against time step of the simulation.
These plots reveal that for large values of the parameter $r$ the transition from the unbalanced (high noise) to the balanced (low noise) phase is more and more sharp. 
Further, in \Cref{fig:UvsT,fig:UvsT_M=1024} a series of plots of work function $U(T)$ are shown for various system size $M$ and the range of interaction $r$. 
As it is seen in these plots, $U=-1$ below some critical value $T_c$ and it gets to zero above $T_c$. 
The transition is apparently more sharp for larger range $r$ of coupling between vertices. 
Also, $T_c$ does not depend on the system size $M$, but it increases with the range $r$. 
This dependence $T_c(r)$ is shown in \Cref{fig:Tcvsr}. 

\section{Discussion}
The spectra of energy in the stationary stage of the time evolution contain not only single states but also blinking states with limit cycles of length two, four, eight and perhaps longer for larger systems. When the system increases with fixed $r$, the width of the spectra are more narrow; note the logarithmic scale of the vertical axis in \Cref{fig:r2histoprob,fig:r3histoprob,fig:r4histoprob}. 
Consequently, the probability of the states with $U=-1$ decreases with the system size $M$. 
For larger systems, practically all states are blinking. 
The mean value of energy, calculated from the spectra, does not change with the system size $M$ and it decreases with the range $r$ of interaction.

On the contrary to the deterministic algorithm reported in the \Cref{sec:deterministic}, the system in the presence of a weak noise is able to get the balance.
The transition point $T_c$ above which the balance vanishes does not depend on the system size. 
Apparently what is relevant for the transition point is the mean degree of vertices. 
In this sense the transition between balanced and non-balanced phase in the chain is more akin to classical
phase transitions than the same transition in complete graphs \cite{PhysRevE.100.022303} or the ferro-paramagnetic transition in scale-free networks \cite{Aleksiejuk_2002}. If our results apply  also for random networks, the transition point there should increase with the mean degree of a node, similarly as we have observed for the chain. On the other hand, the transition presented here is more and more sharp with increasing $r$. This result suggests that the transition on a random network could also sharpen with mean degree of nodes. 

It is worthwhile to note that our result that $T_c$ does not change with $M$ appears to be intermediate when compared to literature. In Ref. \cite{1911.13048} the critical temperature has been shown to increase with the system size, while in Ref. \cite{PhysRevE.100.022303} it decreases to zero in the thermodynamic limit. To resolve this puzzle deserves a more complete study.

We note that for one triad, the condition for HB can be translated into relations between three Boolean variables: if positive is TRUE and negative is FALSE,
then each link is updated as XNOR of the remaining two. The phase transition reported here links the Heider balance to a more wide
set of problems on stability of Boolean networks in the presence of noise \cite{Goodrich_2007,Serra_2010}. Our results add to the area of research on conditions, where a noise allows to get into a global minimum of work function but is not as strong as to remove the system from there (see Refs. \cite{PhysRevE.75.045101,Biondo_2013,PhysRevE.87.022910,Shirado_2017,2002.05451} for examples ranging from the games theory to simulations of opinion formation). 

\bibliographystyle{elsarticle-num}
\bibliography{heider,this,km}

\end{document}